# NOTE ON MINIMALLY $k$-CONNECTED GRAPHS

R. Rama[a], Suresh Badarla[a]

[a] Department of Mathematics, Indian Institute of Technology, Chennai, India

**ABSTRACT**

A $k$-tree is either a complete graph on ($k$+1) vertices or given a $k$-tree $G'$ with $n$ vertices, a $k$-tree $G$ with ($n$+1) vertices can be constructed by introducing a new vertex $v$ and picking a $k$-clique $Q$ in $G'$ and then joining each vertex $u$ in $Q$. A graph $G$ is $k$-edge-connected if the graph remains connected even after deleting fewer edges than $k$ from the graph. A $k$-edge-connected graph $G$ is said to be minimally $k$-connected if $G \setminus \{e\}$ is no longer $k$-edge-connected for any edge $e$ belongs to $E(G)$ where $E(G)$ denotes the set of edges of $G$. In this paper we find two separate $O(n^2)$ algorithms so that a minimally 2-connected graph can be obtained from a 2-tree and a minimally $k$-connected graph can be obtained from a $k$-tree. In a $k$-tree ($k \geq 2$) we find the edges which are insensitive to the $k$-connectivity have both their end vertices of degrees greater than or equal to $k$+1. This property is fully exploited to find an algorithm which reduces any $k$-tree to a minimally $k$-connected graph.

1. **INTRODUCTION**

A chord of a cycle $C$ in a graph $G$ is an edge of $G$ not in $C$ but has an ends in $C$. A chord less cycle in a graph $G$ is a cycle of length greater than equal to four and has no chord. A graph is said to be Chordal iff it has no chord less cycle. That is, every cycle of length greater than four has at least one chord. Chordal graphs are also known as triangulated graphs. $k$-trees are a subclass of a triangulated graphs [1] and have applications in the study of linear sparse systems, scheduling and relational database systems [2]. There are several areas where cliques have very high relevance. We believe that $k$-trees and minimally $k$-connected graphs also have a significant role in the field of Protein Network.

We refer [3] for standard notations and definitions. A $k$-clique is a complete graph on $k$ vertices. A subset $F$ of the edge set $E(G)$ of a graph $G$ is said to be a disconnecting (edge) set iff $G \setminus F$ is disconnected. i.e. $G \setminus F$ has more than one component. Let $k$ be a positive integer. A graph $G$ is said to be $k$-edge-connected iff every disconnecting edge set of G has at least $k$ edges. The concept of recursive labeling, introduced in [4], can be applied to $k$-trees, as we can see in the next definition. A $k$-tree $G$ having n vertices, $n \geq k$ labeled 1, 2,…, $n$ is recursively labeled if vertices 1, 2,…, $k$ form a $k$-clique in $G$, and every vertex labeled $j$, $j > k$, is adjacent to exactly $k$ vertices labeled with values smaller than $j$. The first $k$ vertices of the $k$-tree are called the base-clique. A vertex of $G$ is simplicial if its neighborhood in $G$ is a clique. The degree of a vertex $v$ in $G$, denoted by $d_G(v)$ is the number of edges passing through $v$, a loop being counted twice. A cycle with $n$ vertices is denoted by $C_n$. Let     be the family of $k$-trees. In the following sections, we study some properties on connectivity of 2-trees and $k$-trees. Using these properties we obtain two different algorithms one for 2-tree and one for $k$-tree which brings a 2-tree or a $k$-tree to be a minimally 2-connected or $k$-connected graphs respectively [5]. Here after a 2-connected ($k$-connected) graph means 2 ($k$)-edge-connected graph.

## 2. PROPERTIES OF 2-TREES AND $k$-TREES ($k \geq 2$)

Throughout this section we consider $k$-tree with at least ($k+2$) vertices ($k \geq 2$).

**Lemma 2.1**
Let $G$ be a 2-tree on $n$ vertices and $F$ be the collection of edges present in more than one 3-clique in $G$ then $G \setminus F = G'$ is a minimally 2-connected graph.

**Proof:** We prove the lemma by induction on number of vertices $n$.
Base case: Consider a 2-tree on 4 vertices. It has only one edge $e$ common to both 3-cliques in the 2-tree. Now $G \setminus e = G'$ is minimally 2-connected graph. Hence the claim holds for the base case.
Hypothesis: We assume that the claim holds for a 2-tree on $n$ vertices. This implies that in an $n$ vertex 2-tree $G$, by deleting all the edges present in more than one 3-clique in the 2-tree, results in a minimally 2-connected graph $G'$. Let $F$ be the set of edges that removed from $G$ to form $G'$. Let us assume that there is an edge $\{v_i, v_j\}$ between the vertices in $G'$. This implies that the edge $\{v_i, v_j\}$ is a part of only one 3-clique in $G$ because, if edge $\{v_i, v_j\}$ was part of more than one 3-clique in $G$, it would be removed from $G$. Rest of the edges present in $G'$ are also part of only one 3-clique in $G$.
Inductive step: We now prove that the claim holds for a 2-tree on ($n+1$) vertices. Let $G''$ be a 2-tree on ($n+1$) vertices. Let $v$ be any simplicial vertex of $G''$ and $v$ be adjacent to $v_i, v_j$ of $G''$.
Case-1: If $\{v_i, v_j\}$ is an edge present as common edge to only two 3-cliques of $G''$, then consider the graph $G$ obtained from $G''$ by removing $v$ from it. Now $G$ is a 2-tree on $n$ vertices. By induction, we know that a minimally 2-connected graph $G'$ can be obtained from $G$. Now add vertex $v$ adjacent to the vertices $v_i, v_j$. The resultant graph has exactly one 3-clique which is ($v, v_i, v_j$). Remove the edge $\{v_i, v_j\}$, which gives minimally 2-connected graph on ($n+1$) vertices.
Case-2: If $\{v_i, v_j\}$ is an edge present in more than two 3-cliques, then removal of the simplicial vertex $v$, reduces the number of 3-cliques by one. Hence the graph $G$ thus obtained is a 2-tree on $n$ vertices. Using induction step, a minimally 2-connected graph $G'$ on $n$ vertices can be obtained from $G$. In $G'$, $v_i$ and $v_j$ are not adjacent. Add vertex $v$ to $G'$ by making adjacent to $v_i$ and $v_j$, thus producing a minimally 2-connected graph on ($n+1$) vertices.

**Definition:**
An edge $e$ of $E(G)$ in a $k$-connected graph $G(V, E)$ is insensitive if $G \setminus e$ is $k$-connected.

**Lemma 2.2**
Let  be a family of 2-trees. Then there exists exactly one 2-tree G on $n$ vertices such that $G \setminus F$, where $|F| = 1$ is a minimally 2-connected graph.

**Proof:** A base-clique in a 2-tree will be a 2-clique which is nothing but an edge. Hence a common base- clique in a 2-tree refers to an edge which may be present in two or more 3-cliques. If we assume that $G$ has a single common base clique which is common for all 3-cliques present in $G$, then the common base-clique is edge $e_i$. This implies that $e_i$ is present in all 3-cliques of the 2-tree. Now $G \setminus \{e_i\} = G'$ where $G'$ is minimally 2-connected graph. Hence $|F| = 1$ when $G$ has an only one base-clique.

**Lemma 2.3** Let  be a family of 2-trees. Then there exists a 2-tree on $n$ vertices such that $G \setminus F$ is minimally 2-connected graph which is also a cycle $C_n$, then $|F| = n - 3$.

**Proof:** We give an inductive method of construction of such a 2-tree.
Base Case: Consider a 2-tree on 4 vertices. On removal of the common base-clique which is nothing but an edge present in both the 3-cliques in the graph, we get a minimally 2-connected graph and it is a cycle, $C_4$. Here $n = 4$, $|F| = 1$. i.e. $|F| = 4 - 3 = 1$. Thus the statement holds for base case.
Hypothesis: We assume that the claim holds for a 2-tree on $n$ vertices. That is $G$ be a 2-tree on $n$ vertices and $G \setminus \{F\}$ is minimally 2-connected which is a cycle $C_n$ and $|F| = n - 3$.
Inductive step: We now prove that the claim holds for a 2-tree on $(n+1)$ vertices. That is we have to show that $|F'| = (n+1) - 3$, where $|F'|$ is the number of edges whose removal from the 2-tree of $(n+1)$ vertices leads to a minimally 2-connected graph. For the construction of a 2-tree on $(n+1)$ vertices we use $G$ and introduce a new vertex $v$ adjacent to the vertices $v_i$ and $v_k$ such that $v_i$ and $v_k$ are adjacent and present in exactly one 3-clique in $G$. Let this new graph be $G_1$. If we remove $F$ from $G_1$ we are left with a new graph $G_2$ is $C_n \cup (v, v_i, v_k)$. Now $G_2$ is not minimally 2-connected and removal of edge $\{v_i, v_k\}$, leads to a minimally 2-connected graph $C_{n+1}$. Hence for the additional vertex that is added to form a new cycle $C_{n+1}$, two edges are added and one edge is removed. Therefore, if $|F'|$ is the number of edges we need to remove from 2-tree on $(n+1)$ vertices to form minimally 2-connected graph $C_{n+1}$, then $|F'| = |F| + 2 - 1$. We know that $|F| = n - 3$ from hypothesis. Therefore $|F'| = n - 3 + 2 - 1 = (n+1) - 3$. This completes the induction and the lemma is proved.

**Lemma 2.4**
If $G$ is a 2-tree on $n$ vertices and $G \setminus F = G'$, where $G'$ is a minimally 2-connected graph and $|F| = n - 3$, then $G'$ is a cycle $C_n$.

**Proof:** Proof is similar to the Lemma 2.3.

**Theorem 2.1**
If $G$ is a 2-tree on $n$ vertices and $G \setminus F = G'$, where $G'$ is a minimally 2-connected, then $1 \leq |F| \leq n - 3$.

**Proof:** The proof is immediate consequence of Lemma 2.2 and Lemma 2.3.

Now we give some results on $k$-trees.

**Lemma 2.5**
If $G$ is a k-tree on $n$ vertices and $\{v_i, v_j\}$ is an edge of $G$ whose end vertices having degree greater than or equal to $k+1$ then $G \setminus \{v_i, v_j\}$ is k-connected graph.

**Proof:** Let $G$ be a $k$-tree on $n$ vertices and $\{v_i, v_j\}$ is an edge of $G$ such that $d_G(v_i)$ and $d_G(v_j)$ $\geq k+1$. By the definition of $k$-tree, the degree of each vertex is at least $k$. Then there are more than $k$ disjoint trails from the vertex $v_i$ to vertex $v_j$. Hence $\{v_i, v_j\}$ is an insensitive edge. Hence the proof.

**Theorem 2.2**
If $G$ is a minimally $k$-connected graph obtained from a $k$-tree, then there does not exists an edge $\{v_i, v_j\}$ of $G$ such that $d_G(v_i)$ and $d_G(v_j)$ are greater than or equal to $k+1$.

**Proof:** Let $G$ be a minimally $k$-connected graph obtained from a $k$-tree and let $e = \{v_i, v_j\}$ be an edge of $G$ such that $d_G(v_i), d_G(v_j) \geq k+1$. Consider $G \setminus e$. In the resultant graph there will be exactly $k$ disjoint trails between $v_i$ and $v_j$. Hence $e = \{v_i, v_j\}$ becomes an insensitive edge which implied $G \setminus e$ is minimally $k$-connected. This is a contradiction.

**Remark**
Lemma 2.5 is not true for any graph $G$.

**Preposition 2.1**
Let $G$ be a $k$-tree with $n$ vertices. Then there are $n - k$ cliques with size $(k+1)$ in $G$

**Proof:** The proof of the proposition 2.1 follows from recursive construction of $k$-tree [4]

## 3. ALGORITHMS FOR MINIMALLY 2-($k$)-CONNECTED GRAPHS

In this section we give two different algorithms for reducing a 2-tree ($k$-tree) to a minimally 2($k$)-connected graphs. We establish the correctness and discuss complexity.

**Algorithm 3.1**

Input:   2-tree $G(V, E)$
Output:  A minimally 2-conncted graph $G'(V, E')$
Step1   Find out all the 3-cliques present in the graph
Step2   Let $\{e_1, e_2, ..., e_t\}$ be the edges present in more than one 3-clique
Step3   for $i = 1$ to $t$
            $G \leftarrow G \setminus e_i$

The running time of the algorithm 3.1 is $O(n^2)$. In step1, finding 3-cliques present in the graph will be $O(n^2)$ time, where as step2 and step3 will take constant time for inspecting the edges and removing the edges from the given input graph. The correctness of the algorithm 3.1 follows from Lemma 2.1.

**Algorithm 3.2**

Input:   $k$-tree $G(V, E)$
Output:  minimally $k$-connected graph $G'(V, E')$

Step1   Label the vertices of the input graph in the increasing order of degree sequence
        LIST ←
Step2   Let $\{v_i,...,v_n\}$ are the vertices of degree $\geq (k+1)$
        *for* s = i to n
        *for* t = i+1 to n
        *if* ($\{v_s, v_t\} \in E(G)$) *then*
        LIST ← $\{v_s, v_t\}$
Step3   *for* p = i to n
        *for* q = i+1 to n
        *if* (($d_G(v_p)$ & $d_G(v_q)$) $\geq (k+1)$ & $\{v_p, v_q\} \in$ LIST)
        *then* $G \leftarrow G \setminus \{v_p, v_q\}$
        LIST ← LIST $\setminus \{v_p, v_q\}$
        *else*
        continue;

The running time of the algorithm 3.2 will be $O(n^2)$. Sorting the vertices of the input graph in the ascending order of degree sequence will take the linear time where as step2 and step3 will take $O(n^2)$ time. Therefore the total running time of the algorithm 3.2 will be $O(n^2)$. Justification for correctness of algorithm of 3.2 is as follows. Since the connectivity of a graph is dependent on the degree of vertices, in step1, we label the vertices of input graph according to the increasing sequence of degree. In step2, we are listing the edges whose end vertices having degree greater than or equal to $k+1$. In step3, we start with the vertices of degree greater than or equal to $k+1$, since by Lemma 2.5 the insensitive edges will have their end degrees greater than or equal to $k+1$. So every time we are going to remove the edges whose adjacent vertices are of degree greater than or equal to $k+1$ which are insensitive by Lemma 2.5. After removal of some edges according to the *for* loop, some of the vertices may have degree greater than or equal to $k+1$. These vertices are non adjacent because if they are adjacent the conditions in *if* statement are true and that edge would have been removed from the graph. So the vertices of degree greater than or equal to $k+1$ are adjacent only with the vertices whose degree equal to $k$. If we remove any edge from the obtained graph, the graph is not $k$-connected by theorem 2.2. So finally we are able to get minimally $k$-connected graph.

## 4. CONCLUSION AND FUTURE SCOPE

We have presented an algorithm using which we can obtain a minimally 2-connected graph from a given 2-tree. On a similar vein, we have presented an algorithm using which we can obtain a minimally $k$-connected graph from a given $k$-tree. We have also found out the kind of edges that are needed to be removed from a $2(k)$-tree to form a minimally $2(k)$-connected graph.

Future scope revolves around 'find the class/classes of minimally $k$-connected graphs which can be obtained from $k$-trees'. Some characterizations are given [6]. Is it possible to construct a $k$-tree from a minimally $k$-connected graph which is also a partial $k$-tree? We would also like to know whether the decision problem "Is a minimally $k$-connected graph a partial $k$-tree (an edge sub graph of $k$-tree)?" is an NP-hard or not.